\documentclass{aa}

\usepackage{graphicx}
\usepackage{txfonts}
\usepackage{natbib}
\usepackage[figuresright]{rotating}
\usepackage{lscape}
\usepackage{aalongtable}
\usepackage{afterpage}

\begin{document}

   \title{A spectroscopic study of the near-IR [SIII] lines in a sample of HII galaxies: chemical abundances}

   \author{C. Kehrig
          \inst{1,2}
          \and
          J.M. V\'{\i}lchez\inst{1}
          \and
          E. Telles\inst{2}
          \and
          F. Cuisinier\inst{3}
          \and
          E.P\'erez-Montero\inst{4}
          }

   \offprints{C. Kehrig}

   \institute{Instituto de Astrof\'{\i}sica de Andaluc\'{\i}a (CSIC),
              Apartado 3004, 18080 Granada, Spain\\
              \email{kehrig@iaa.es,jvm@iaa.es}
         \and
             Observat\'{o}rio Nacional,
             Rua Jos\'{e} Cristino, 77, 20.921-400, Rio de Janeiro - RJ, Brazil\\
             \email{kehrig@on.br,etelles@on.br}
          \and
             GEMAC, Observat\'{o}rio do Valongo/UFRJ,
             Ladeira do Pedro Ant\^{o}nio, 43, 20.080-090 Rio de Janeiro - RJ, Brazil\\
             \email{francois@ov.ufrj.br}
          \and
             Departamento de F\'{\i}sica Te\'orica,C-XI,Universidad Aut\'onoma de Madrid, 28049 Madrid, Spain\\
             \email{enrique.perez@uam.es}
              }

   \date{Received <date>; accepted <date>; Last update \today}
   
  \abstract
   \textbf{{A detailed spectroscopic study covering the blue to near-infrared
   wavelength range ($\lambda$3700~\AA~-1$\mu$m) was performed for a
   sample of 34 HII galaxies in order to derive fundamental parameters
   for their HII regions and ionizing sources, as well as gaseous
   metal abundances.  All the spectra included the nebular
   [SIII]$\lambda$$\lambda$9069,9532~\AA~lines, given their importance
   in the derivation of the S/H abundance and relevant ionization
   diagnostics.}}
   {A systematic method was followed to correct the near-IR [SIII]
   line fluxes for the effects of the atmospheric transmission. A
   comparative analysis of the predictions of the empirical abundance
   indicators R$_{23}$ and S$_{23}$ was performed for our sample
   galaxies.  The relative hardness of their ionizing sources was
   studied using the $\eta$' parameter and exploring the role played
   by metallicity.}
   {For 22 galaxies of the sample, a value of the electron temperature
   T$_{e}$[SIII] was derived, along with their ionic and total S/H
   abundances. Their ionic and total O/H abundances were derived using
   direct determinations of T$_{e}$[OIII]. For the rest of the
   objects, the total S/H abundance was derived using the S$_{23}$
   calibration. The abundance range covered by our sample goes from
   1/20 solar up to solar metallicity. Six galaxies present 12+log
   (O/H) $<$ 7.8 dex. The mean S/O ratio derived in this work is log
   (S/O)=-1.68$\pm$0.20 dex, 1$\sigma$ below the solar (S/O)$_\odot$
   value. The S/O abundance ratio shows no significant trend with O/H
   over the range of abundance covered in this work, in agreement with
   previous findings. There is a trend for HII galaxies with lower
   gaseous metallicity to present harder ionizing spectra. We compared
   the distribution of the ionic ratios O$^{+}$/O$^{++}$ vs.
   S$^{+}$/S$^{++}$ derived for our sample with the predictions of a
   grid of photoionization models performed for three different
   stellar effective temperatures. This analysis indicates that a
   large fraction of galaxies in our sample seem to be ionized by
   extremely hard spectra, in line with recent suggestions for extra
   ionizing sources in HII galaxies.}
   {}


   \maketitle

   \section{Introduction}
   \label{intro}
   
   HII galaxies are galaxies undergoing violent star
   formation (Searle \& Sargent 1972; Terlevich et al. 1991; Cair\'os
   et al. 2000). Their optical spectra show strong emission lines
   (recombination lines of hydrogen and helium, as well as forbidden
   lines of elements like oxygen, neon, nitrogen, sulfur, among
   others) that are very similar to the spectra of extragalactic HII
   regions. Analysis of their spectra shows that they are
   low-metallicity objects with the metallicity varying from
   1/40$Z_{\odot}$ to 1/2$Z_{\odot}$ (e.g. Terlevich et al. 1991;
   Telles 1995 and references therein; V\'{\i}lchez \&
   Iglesias-P\'aramo 1998,2003; Thuan \& Izotov 2005). Among them we
   can find the least chemically-evolved galaxies in the local
   Universe.

   The study of elemental abundances in emission-line galaxies gives
   information about their chemical evolution and star formation
   history. Outside the Local Group, emission lines from ionized gas
   represent the principal means of deriving abundances, as energy is
   concentrated in a few conspicuous emission lines. Abundances for
   the stellar population are derived from absorption features, which
   are more numerous and require much higher signal-to-noise spectra
   to be derived meaningfully.

    In HII galaxies the metal enrichment of the interstellar medium by
    supernovae has been operating typically in low-metallicity
    environments.  Oxygen is the most frequently used element in
    deriving abundances from emission lines: abundances are easily
    derived, as the main ionization stages are observable in the
    optical range. Furthermore, oxygen is particularly suitable for
    chemical evolution studies, as it traces the overall metallicity
    very well. It originates quasi exclusively from the
    nucleosynthesis in type II supernovae progenitors (Meynet \&
    Maeder 2002; Pagel 1997; Woosley \& Weaver 1995). While the
    sources of oxygen are well-determined and the most important
    ionization stages can be observed in the optical range, some uncertainties still
    remain  about the sulfur yields and its
    sources. In addition, not all the ionization stages can be
    observed in the optical range and important ionization correction
    factors (ICFs) must be applied to derive the total sulfur
    abundance. Hence comparing S and O abundances can
    give us some clues to sulfur nucleosynthesis and the masses of
    the stars where the sulfur tends to be formed.

   To derive oxygen abundances, one should first derive the
   electron temperature, which requires the measurement of faint
   auroral lines, like [OIII]$\lambda$4363~\AA, which are often not
   detected. The alternative is to use strong line-abundance
   indicators, like R$_{23}$\footnote{R$_{23}$=([OII]$\lambda$3727 +
   [OIII]$\lambda$$\lambda$4959,5007)/H$\beta$}, which calibrated
   empirically (Pagel et al. 1979, Pilyugin 2001) or through
   photoionization models (e.g. McGaugh 1991). However, the relation
   between R$_{23}$ and oxygen abundance presents the noticeable
   drawback of being double-valued.

  V\'{\i}lchez \& Esteban (1996) proposed S$_{23}$\footnote{S$_{23}$ =
  ([SII]$\lambda$$\lambda$6717,31 + [SIII]$\lambda$$\lambda$9069,9532)/H$\beta$} as an
  alternative abundance indicator. In contrast to oxygen, S$_{23}$
  remains single-valued up to abundances above solar
  value. Furthermore, sulfur should be as useful as oxygen for tracing
  metallicity. From an observational point of view, S$_{23}$ has the
  advantage over R$_{23}$ that the [SII] and [SIII] lines are less
  affected by reddening (P\'erez-Montero et al. 2005; hereinafter
  PM06).

   To produce an accurate derivation of S/H
   abundance, the importance of using the nebular [SIII] lines can not
   be overlooked (e.g. Dennefeld \& Stasi\'nska 1983; V\'{\i}lchez et
   al. 1988, Garnett 1989; Bresolin et al. 2004). Photoionization models
   indicate that S$^{++}$ is the dominant sulfur ion (Garnett 1989; hereinafter G89), which
   presents three forbidden transitions at [SIII]$\lambda$$\lambda$9069,9532~$\AA$~
   and $\lambda$6312~$\AA$~ in the optical to near-IR (NIR) range (analogs
   to [OIII]$\lambda$$\lambda$4959,5007~$\AA$~ and
   $\lambda$4363~$\AA$). The [SIII]$\lambda$6312~$\AA$~line is
   faint, highly temperature-sensitive, and it can induce several biases in
   the derived S/H abundance. The NIR [SIII] lines can be quite
   strong, and a detailed telluric atmosphere correction has to be
   applied to them. P\'erez-Montero \& D\'{\i}az
   (2003) (hereinafter PMD03) and G89 derive the S$^{++}$ ionic
   abundance for samples of about one dozen emission-line galaxies,
   both using the nebular [SIII]$\lambda$9069~$\AA$~line. Recent work
   by Izotov et al. (2005) (hereinafter I05) presents S/H abundances
   for a large number of metal-poor emission-line galaxies from the
   SDSS-DR3\footnote{Data Release 3 of Sloan Digital Sky Survey};
   however, the auroral line [SIII]$\lambda$6312~$\AA$~
   was used in this work to calculate the S$^{++}$ ionic abundance.

   Here we present long-slit spectrophotometric
   observations of a sample of 34 HII galaxies to make a
   detailed analysis of their chemical abundances. The wide coverage
   of our spectra ($\lambda$3700~\AA~ - 1$\mu$m) for all the galaxies
   in the sample provides us all the emission lines needed to
   estimate the oxygen and sulfur abundances directly. All the S/H
   abundances were estimated using a nebular [SIII] line, so that
   uncertainties related to the use of the auroral line
   [SIII]$\lambda$6312~\AA~are avoided.  In addition, this wavelength
   coverage allowed us to study the properties of the ionizing
   clusters of HII galaxies making use of the $\eta$' parameter and
   sequences of photoionization models.

  In the next section we describe our sample of galaxies, the
  observations, and data reduction and present the line
  intensities. In Sect.3 we perform a comparative study between
  R$_{23}$ and S$_{23}$ abundance indicators, present an analysis
  about ionization structure and ionizing sources, and discuss the
  abundance results for the sample. Finally in Sect.4 we summarize
  our conclusions.
\section{Data analysis}

\subsection{Sample and observations}
\label{spectra}

   The data base of this work consists of 34 intermediate-resolution
   spectra of HII galaxies covering a wavelength range from
   3700~\AA~to 7000~\AA~ (blue spectra; Kehrig et al. 2004), and from
   6500~\AA~ to 1$\mu$m (red spectra). For all the objects,
   measurements of the emission lines of [OII]$\lambda$3727~\AA~and
   [SIII] $\lambda\lambda$9069,9532~\AA~ exist, except for the galaxy
   UM151, for which we do not have a measurement of the
   [OII]$\lambda$3727~\AA~line.

   The complete log and the characteristics of the sample objects are
   given in Kehrig et al. (2004). The mean value for the distribution
   of redshifts of the sample is 0.02. Depending on the redshift of
   each galaxy, one of the two [SIII] lines ($\lambda$9069~\AA~ or $\lambda$9532~\AA)
   may fall in the range of telluric absorption in which 
   atmospheric correction is critical.  For this reason, this
   correction must be performed on a case by case basis.

   Regarding the red spectra, the observations were conducted in
   October and December 2002 (9 nights of observations in total) with
   the Boller \& Chivens spectrograph of the 1.52 m
   telescope\footnote{operated within the agreement between Brazil and
   ESO} at the European Southern Observatory (ESO), La Silla,
   Chile. The CCD used has a pixel size of 0.82 arcsec in spatial
   direction.  Typical seeing of the observations was 1-1.2$''$.  All
   observations were performed using grating \#10 with an inverse
   dispersion of 1.9~\AA/pix, a slit width of 2.5$''$, and a spectral
   range of 6000~\AA~-1$\mu$m. This configuration yielded an effective
   instrumental resolution of \mbox{$\sim$ 6~\AA~ (FWHM)} at
   6000~\AA. Total exposure times were typically 7200 seconds split into
   two exposures in order to eliminate cosmic rays during the
   reduction procedure.

\begin{figure}
   \includegraphics[width=\columnwidth]{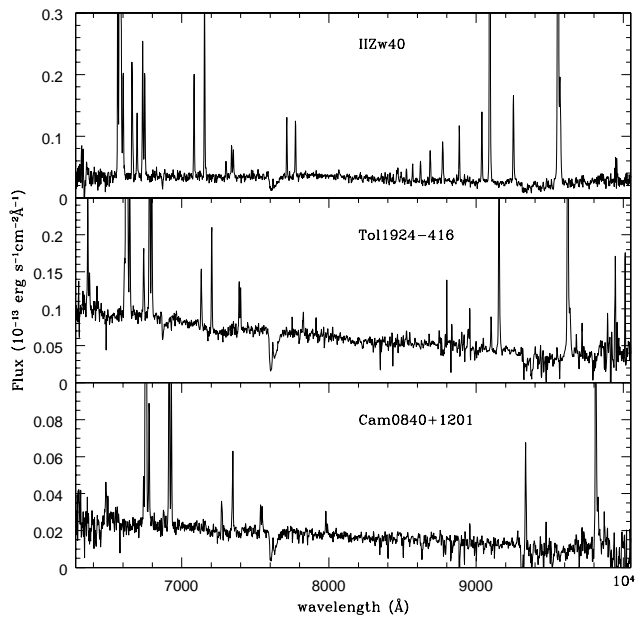}
      \caption{Representative red spectra of galaxies of the sample.
              }
         \label{spectra}
   \end{figure}

Representative red spectra of three of the observed galaxies are shown in
Fig.~\ref{spectra}.

\subsection{Data reduction and telluric absorption correction}

  The CCD frames were reduced by employing standard IRAF\footnote{IRAF
  is distributed by the National Optical Astronomy Observatories}
  packages.  The spectrophotometric standard stars ($\sim$ 7 observed
  each night) used for flux calibration were chosen in order to have
  an appropriate flux-point coverage in the NIR.
  
Ground-based NIR spectroscopy has always been hampered by strong and
variable absorption features due to the Earth's atmosphere. Even
within the well-established photometric bands such as J, H, and K,
telluric absorption bands are present. In analysis of the NIR sulfur
emission lines, a crucial step is the correction for the effects
produced by the earth's atmosphere on the spectra, especially between
8500~\AA~and 1$\mu$m (Vacca et al. 2003; D\'{\i}az et
al. 1987). Exhaustive work was done to correct the whole sample for
these effects. We derived, for each night, the telluric correction as
a function of wavelength, $<$f($\lambda$)$>$, and its corresponding
standard deviation ,$<$$\sigma$(f($\lambda$))$>$.  A minimum of five
standard stars per night was used to obtain this correction. The mean
values for $<$f($\lambda$)$>$ and $<$$\sigma$(f($\lambda$))$>$ are
80$\%$ and 4$\%$, respectively.  We applied the telluric correction by
dividing each galaxy spectrum by its corresponding
$<$f($\lambda$)$>$. The $<$ $\sigma$(f($\lambda$))$>$ of the
correction was taken into account when calculating the overall error
budget of the line fluxes.

\subsection{Line intensities}

\begin{sidewaystable*}
\centering
\renewcommand{\footnoterule}{}  
\caption{\label{fluxes} Reddening corrected line fluxes, relative to H$\beta$=100, and corresponding extinction coefficients for the sample of galaxies.}
\tiny{
\begin{tabular}{lcccccccccccccccccccc}
\hline
\hline

Galaxy & [OI]  & [SIII] & [OI]   & [NII] &  H$\alpha$ & [NII]  & HeI   & [SII] & [SII]  & HeI  & [ArIII] & [OII]  &[OII] & Pa12 & Pa11   &  Pa10  &  Pa9  & [SIII] & Pa8 &C(H$\beta$) \\

& 6300  & 6312 & 6364   & 6548  & 6563   & 6584   & 6678  & 6717  &  6731  & 7065 &  7136   & 7320   & 7330 &8750  &8865         &9014        &9229       & 9069      &9548  & \\

(1)     &  (2)     &  (3)     & (4)     &  (5)     &  (6)     & (7)     &  (8)     &  (9)     & (10)    &  (11)    &  (12)    & (13)    &  (14)    &  (15)    & (16)    &  (17)    &  (18)    & (19)    &  (20) & (21) \\ \hline \\

UM238  &  --- &  --- &  --- &  5.7 &286.0&  8.6&  ---& 17.2& 13.4&  5.8&  6.8&  3.2&  1.8&  ---&  ---&  ---&  ---&  3.7&  4.7 &0.26 \\

 &  ---&  ---&  ---& $\pm$  1.1&$\pm$  5.5  & $\pm$  1.3& ---& $\pm$  1.8& $\pm$  1.9& $\pm$  0.9& $\pm$  0.3& $\pm$  0.7& $\pm$  0.5  &  ---& ---& ---  & ---& $\pm$  0.5& $\pm$  1.0 & \\
 
UM69(E) &  --- & --- &  --- &  9.4 &286.0 & 28.9 &  ---& 40.5& 30.0&  ---&  8.4&  7.2&  3.4&  ---&  ---&  ---&  ---& 16.7& 4.7 &0.74 \\ 

 & ---&  ---&  ---& $\pm$  1.4&$\pm$  2.3  & $\pm$  1.4& ---& $\pm$  1.6& $\pm$  1.4& ---& $\pm$  0.9&  $\pm$  0.7& $\pm$  0.6  & ---& ---& ---  &---& $\pm$  1.2& $\pm$  0.9 & \\ 

UM69(W)  &  ---&  ---&  ---& 15.4&286.0& 39.2&  ---& 56.2& 40.0&  ---&  6.4&  ---&  ---&  ---&  ---&  ---&  ---& 14.7&  3.5 &0.74 \\

& ---& ---& ---& $\pm$  1.4&$\pm$  1.6  & $\pm$  1.2& ---& $\pm$  2.2& $\pm$  2.0& ---& $\pm$  0.5& ---& ---  & ---&  ---& ---  & ---& $\pm$  1.0& $\pm$  0.7 & \\ 

Tol0104-388   &  7.1&  ---&  ---&  8.3 &286.0& 22.1&  3.3& 29.7& 21.9&  ---&  9.0&  3.6&  2.8&  ---&  2.0&  1.9&  ---& 48.8$^{a}$ &  4.0 &0.18 \\

&  $\pm$  0.2& ---&  ---& $\pm$  0.5& $\pm$  0.6  & $\pm$  0.2& $\pm$  0.3& $\pm$  0.3& $\pm$  0.1& ---&  $\pm$  0.4& $\pm$  0.1& $\pm$  0.1  &  ---&  $\pm$  0.3& $\pm$  0.3  &  ---& $\pm$  6.5&  $\pm$  0.9 & \\

UM306  &  ---&  ---&  ---&  6.1&286.0& 16.3&  ---& 23.1& 17.9&  ---&  7.0&  ---&  ---&  ---&  ---&  ---&  ---& 22.7&  3.7 &0.21 \\ 

 &  ---& ---& ---& $\pm$  0.9&$\pm$  1.3  & $\pm$  0.7& ---& $\pm$  1.3& $\pm$  1.2& ---& $\pm$  0.9& ---& ---  & ---& ---& ---  & ---& $\pm$  6.0& $\pm$  2.1 & \\ 

UM307  &  ---&  ---&  ---& 15.4&286.0& 46.5&  4.3& 24.0& 27.3&  ---&  4.8&  2.4&  1.5&  ---&  ---&  2.1&  ---& 18.5&  3.7 &0.25 \\ 

 &  ---&  ---&  ---& $\pm$  0.5&$\pm$  1.0  & $\pm$  0.7& $\pm$  1.0& $\pm$  0.7& $\pm$  0.6& ---& $\pm$  0.7& $\pm$  0.4& $\pm$  0.4  & ---& ---& $\pm$  0.7  & ---& $\pm$  1.7& $\pm$  0.6 & \\

Tol0117-414NS(S) &  ---&  ---&  ---& 21.3&286.0 &35.6 & ---&40.3 &31.1 & --- & ---& --- & --- & --- & --- & ---& --- &32.5 & --- &0.10 \\ 

 &  ---&  ---&  ---& $\pm$ 1.8 &$\pm$  2.6  &$\pm$  1.8& ---&$\pm$ 1.6 &$\pm$  1.6 & --- & --- & --- & --- & --- & --- & --- & --- &$\pm$ 12.0 & --- & \\ 

Tol0117-414NS(N)&  ---&  ---&  ---& 20.2&286.0& 36.9&  ---& 43.0& 31.5&  ---&  ---&  ---&  ---& ---&  ---&  ---& ---& 31.7&  --- &0.22 \\ 

&  ---&  ---&  ---& $\pm$  2.3 &$\pm$  6.8 & $\pm$  1.4&  ---& $\pm$  2.0& $\pm$  0.9&  ---&  ---&  ---&  ---  & ---&  ---&  --- & ---& $\pm$ 12.4&  --- & \\

Tol0117-414NS&  ---&  ---&  ---& 20.0&286.0& 38.1 &  ---& 42.9& 32.0&  ---&  ---&  ---&  ---& ---&  ---&  ---& ---& 27.9&  --- &0.10 \\

&  ---&  ---&  ---& $\pm$  2.4&$\pm$  2.9 & $\pm$  1.4&  ---& $\pm$  0.5& $\pm$  1.1&  ---&  ---& ---&  ---  & ---&  ---& ---  & ---& $\pm$  9.6& --- & \\
 
Tol0117-414EW(EW) & 12.1&  ---&  ---& 17.8&286.0& 50.1&  ---& 58.5& 43.3&  ---&  ---&  ---&  ---&  ---&  ---&  ---&  ---& 28.2&  --- &0.31  \\

 & $\pm$  1.1& ---&  ---& $\pm$  1.0&$\pm$  1.9  & $\pm$  1.2&  ---& $\pm$  1.7& $\pm$  2.9& --- & ---& ---& --- & ---&  ---& ---  &  ---& $\pm$ 10.5&  --- & \\

Tol0117-414EW & 16.1 &  --- &  --- &  7.8 &286.0 & 39.2 &  --- & 57.8 & 45.1 &  --- &  ---& ---  &  ---& ---&  --- &  --- & ---& 25.9&  --- &0.31 \\

& $\pm$  1.3& ---  &  ---& $\pm$  0.3&$\pm$  6.6 & $\pm$  1.0&  ---& $\pm$  1.2& $\pm$  1.4&  ---&  ---&  ---& --- & ---& ---& --- & ---& $\pm$  9.0& --- &  \\

UM323   &  --- &  --- &  ---&  9.9&286.0& 17.3&  ---& 26.9& 20.1&  ---&  ---&  ---&  ---&  ---&  ---&  ---&  ---&  9.6&  4.8 &0.88 \\

&  ---& ---& ---& $\pm$  5.1&$\pm$  7.4  & $\pm$  4.7& ---& $\pm$  6.6& $\pm$  5.3& ---& ---&  ---& ---  &  ---&  ---&  ---  & ---& $\pm$  1.3&  $\pm$  1.2 & \\ 

Tol0140-420  &  --- &  --- &  ---&  ---&286.0& 12.4&  ---& 29.7& 20.8&  ---&  ---&  --- &  --- &  ---&  ---&  ---&  ---& 16.5&  --- &0.02 \\ 

&  ---&  ---&  ---&  ---&$\pm$  5.9  & $\pm$  1.0& ---& $\pm$  2.4& $\pm$  2.4&  ---& ---& ---& ---  & ---& ---& ---  & ---& $\pm$  5.4& --- & \\

UM137  &  ---&  ---&  ---&  ---&286.0& 18.1&  ---& 32.1& 28.1&  4.8&  2.2&  ---&  ---&  ---&  ---&  5.1&  ---&  5.3&  4.4  &1.47 \\ 

&  ---&  ---&  ---& ---&$\pm$  2.8  & $\pm$  1.8&  ---& $\pm$  3.2& $\pm$  2.8& $\pm$  1.1& $\pm$  0.7& ---& ---  & ---& ---& $\pm$  1.0  & ---& $\pm$  1.2& $\pm$  1.5 & \\

UM151  &  ---&  ---&  ---& 13.4&286.0& 37.6&  ---& 48.2& 33.4&  ---&  7.7&  ---&  ---&  ---&  ---&  1.9&  ---& 28.5&  2.8 &0.41 \\ 

 &  ---& ---& ---& $\pm$  2.1&$\pm$  3.5  & $\pm$  2.3& ---& $\pm$  3.1& $\pm$  2.5& ---& $\pm$  2.7& ---& ---  &  ---& ---& $\pm$  0.6  & ---& $\pm$  6.4&  $\pm$  0.6 & \\

UM391  &  7.6&  ---&  4.2& 26.2&286.0& 79.6&  ---& 64.5& 25.5&  ---&  5.4&  ---&  ---&  ---&  ---&  ---&  ---& 19.5&  --- &0.53 \\

&  $\pm$  2.0&  ---& $\pm$  1.5& $\pm$  1.3&$\pm$  1.6  & $\pm$  0.8& ---& $\pm$  0.8& $\pm$  0.7&  $\pm$   0.& $\pm$  1.2& ---&     ---& ---& ---&  ---  &  ---& $\pm$  2.9&  --- & \\ 

UM396   &  ---&  ---&  ---&  7.3&286.0&  9.0&  3.1& 13.4&  7.3&  3.7& 10.2&  ---&  ---&  ---&  3.0&  6.7&  ---& 27.3&  3.9 &0.35 \\

&  ---&  ---& ---& $\pm$  1.4&$\pm$  1.9  & $\pm$  1.3& $\pm$  0.4& $\pm$  0.6& $\pm$  0.5& $\pm$  0.5& $\pm$  0.6&  ---&  ---  & ---& $\pm$  0.5& $\pm$  1.0  & ---& $\pm$  5.2& $\pm$  0.3 & \\

UM408  &  ---&  ---&  ---&  ---&286.0&  6.8&  ---& 17.7& 11.5&  ---&  6.9&  ---&  --- &  ---&  ---&  ---&  ---&  5.5 &  --- &1.32 \\

&  ---& ---& ---& ---&$\pm$  4.3  & $\pm$  1.3& ---& $\pm$  1.2& $\pm$  1.1&  ---& $\pm$  0.7& ---& ---  & ---&  ---& ---  & ---& $\pm$  0.8& --- & \\ 

UM417  &  --- &  ---&  ---&  ---&286.0&  8.3&  ---& 10.1&  8.9&  4.5&  4.6&  ---&  ---&  ---&  ---&  ---&  ---&  1.8&  --- &0.47 \\ 

 &  --- &  --- & --- & --- & $\pm$  4.1  & $\pm$  1.3 & --- & $\pm$  0.7 & $\pm$  0.7 & $\pm$  0.7 &  $\pm$  1.2 &  --- & ---  & --- & --- & ---  & --- & $\pm$  0.2 & --- & \\

Tol0226-390  &  9.8&  2.5&  4.1& 15.8&286.0& 31.2&  3.3& 25.9& 21.9&  4.2&  8.8&  3.5&  3.0&  ---&  ---&  ---&  2.1& 20.4&  5.5 &0.34 \\ 

 &  $\pm$  0.3&  $\pm$  0.3&  $\pm$  0.5& $\pm$  0.9&$\pm$  1.3  & $\pm$  1.0&  $\pm$  0.3& $\pm$  0.8& $\pm$  0.7& $\pm$  0.2& $\pm$  0.4& $\pm$  0.3&  $\pm$  0.2  & ---& ---& ---  & $\pm$  0.5& $\pm$  3.4&  $\pm$  2.6 & \\

Tol0306-405    &  --- &  ---&  ---&  ---&286.0 & 14.1&  ---& 24.3& 20.9&  ---&  8.4&  ---&  ---&  ---&  ---&  ---&  ---& 7.5&  --- &0.33 \\ 

&  ---& ---&  ---&  ---&$\pm$  1.6  & $\pm$  0.9&  ---& $\pm$  0.7& $\pm$  0.8&  ---&  $\pm$  0.1& ---& ---  & ---&  ---& ---  &  ---& $\pm$  3.5&  --- & \\

Tol0341-407(W)  &  5.2 &  --- &  --- &  1.1 &286.0 &  6.4 &  4.8 & 15.5 & 10.7 &  2.9 &  6.2&  2.5 &  2.6 &  1.6 &  --- &  --- &  3.0 & 19.1 &  --- &0.06 \\

& $\pm$  0.4& ---& ---& $\pm$  0.1&$\pm$  4.0  & $\pm$  0.1& $\pm$  0.5& $\pm$  0.1& $\pm$  0.1& $\pm$  0.1& $\pm$  0.1& $\pm$  0.2& $\pm$  0.1  & $\pm$  0.5& ---& ---  & $\pm$  0.6& $\pm$  1.7& --- & \\  

Tol0341-407(E)    &  --- &  --- &  --- &  --- &286.0 & 16.6 &  --- & 37.8 & 25.3 &  --- &  --- &  4.8 &  3.8 &  --- &  --- &  --- &  --- & 17.5 &  --- &0.06 \\

  &  ---& ---& ---& ---&$\pm$  4.5  & $\pm$  3.2& ---& $\pm$  3.0& $\pm$  2.4& ---& ---& $\pm$  1.2& $\pm$  1.0  & ---& ---& ---  & ---& $\pm$  2.0& --- & \\ 

Cam0357-3915  &  2.5&  ---&  ---&  2.4&286.0&  6.5&  3.3&  9.3&  6.3&  5.2&  4.8&  1.9&  0.9&  ---&  ---&  2.2&  ---& 11.5&  --- &0.15 \\

& $\pm$  0.2&  ---&  ---& $\pm$  0.1&$\pm$  5.1  &  $\pm$  0.6&  $\pm$  0.2& $\pm$  0.1& $\pm$  0.6&  $\pm$  0.3&  $\pm$  0.1&  $\pm$  0.1& $\pm$  0.1  & ---&  ---&  $\pm$  0.5  &  ---& $\pm$  2.0& --- & \\ 


CTS1006  &  ---&  ---&  ---&  4.9&286.0& 13.7&  3.6& 21.3& 10.8&  3.2&  7.2&  2.4&  1.8&  0.9&  0.9&  1.2&  2.3& 15.2&  5.1 &0.15 \\

&  ---& ---& ---& $\pm$  0.4&---  & $\pm$  0.3& $\pm$  0.4& $\pm$  0.6& $\pm$  0.5& $\pm$  0.4& $\pm$  0.3& $\pm$  0.2& $\pm$  0.2  & $\pm$  0.2& $\pm$  0.1& $\pm$  0.2  & $\pm$  0.5& $\pm$  1.3& $\pm$  0.6 & \\ 

CTS1008 &  3.2 &  1.6&  ---&  5.1&286.0& 13.4&  3.9& 16.7& 13.7&  3.7&  9.5&  3.2&  1.6&  ---&  ---&  ---&  4.2& 17.4&  4.7 &0.24 \\ 

 & $\pm$  0.1& $\pm$  0.4& ---& $\pm$  0.3&$\pm$  0.5  & $\pm$  0.3& $\pm$  0.3& $\pm$  0.4& $\pm$  0.4& $\pm$  0.3& $\pm$  0.6& $\pm$  0.5&  $\pm$  0.3  & ---& ---& ---  & $\pm$  1.0& $\pm$  3.3& $\pm$  1.4 & \\ \hline

\end{tabular}
}
\vfill
\end{sidewaystable*}

\begin{sidewaystable*}
\centering
\renewcommand{\footnoterule}{}  
\addtocounter{table}{-1}
\caption{continued.}
\tiny{
\begin{tabular}{lcccccccccccccccccccc}
\hline
\hline

Galaxy & [OI]  & [SIII] & [OI]   & [NII] &  H$\alpha$ & [NII]  & HeI   & [SII] & [SII]  & HeI  & [ArIII] & [OII]  &[OII] & Pa12 & Pa11   &  Pa10  &  Pa9  & [SIII] & Pa8 &C(H$\beta$) \\

       & 6300  & 6312   & 6364   & 6548  & 6563  & 6584   & 6678  & 6717  &  6731  & 7065 &  7136   & 7320   & 7330 & 8750     &8865         &9014        &9229       & 9069      &9548  & \\

(1)     &  (2)     &  (3)     & (4)     &  (5)     &  (6)     & (7)     &  (8)     &  (9)     & (10)    &  (11)    &  (12)    & (13)    &  (14)    &  (15)    & (16)    &  (17)    &  (18)    & (19)    &  (20) & (21) \\  \hline

Tol0528-383(E) &  --- &  ---&  ---& 10.3 &286.0 & 18.0 &  ---& 39.9 & 30.4& ---& 11.3 &  ---&  ---&  ---&  ---&  ---&  ---& 19.9& 3.9 &0.46 \\ 

 & ---&  ---&  ---& $\pm$  2.9&$\pm$  3.2  & $\pm$  2.9& ---& $\pm$  2.1& $\pm$  2.7& ---& $\pm$  1.9& ---& ---  &  ---&  ---&  ---  & ---& $\pm$  3.0&  $\pm$  1.3  & \\

Tol0528-383(W) &  --- &  ---&  ---& 10.2&286.0& 16.4&  --- & 31.6 & 22.0&  ---&  9.2&  ---&  ---&  ---&  ---&  ---&  2.7& 13.5&  5.0 &0.46 \\

 &  ---&  ---&  ---& $\pm$  1.9&$\pm$  2.0  & $\pm$  1.8& ---& $\pm$  1.8& $\pm$  1.4& ---& $\pm$  1.6& ---& ---  &  ---& ---&  ---  & $\pm$  0.9& $\pm$  2.5&  $\pm$  1.5 & \\

Tol0538-416  &  4.5&  2.0&  3.0&  5.1&286.0   &14.9  & 4.0  &24.4  &15.9  & 3.4  &16.7  & ---& --- & --- & 2.3  & --- & --- &19.2 & 4.7 &0.14 \\ 

  &  $\pm$ 0.4&  $\pm$ 0.3&  $\pm$ 0.4&  $\pm$ 0.7& $\pm$  0.9  &$\pm$  0.6 & $\pm$  0.4 &$\pm$  0.8 & $\pm$0.6 & $\pm$  0.5 & $\pm$ 0.5 & ---& --- & --- & $\pm$ 0.6 & ---  & ---&$\pm$  4.1& $\pm$  2.9 & \\ 

IIZw40  &  1.8&  1.1&  ---&  ---&286.0&  6.4&  3.5 &  7.6&  6.2&  4.6&  8.3&  1.4 &  1.1 &  1.1&  1.3&  1.9&  2.7& 15.3&  4.3 &1.29 \\ 

 &  $\pm$  0.2& $\pm$  0.2& ---&---&$\pm$  0.3  & $\pm$  0.2&$\pm$  0.3& $\pm$  0.4& $\pm$  0.3&$\pm$  0.2& $\pm$  0.3& $\pm$  0.2& $\pm$  0.2  & $\pm$  0.1& $\pm$  0.1& $\pm$  0.3  &  $\pm$  0.3& $\pm$  1.6&  $\pm$  1.4 & \\ 

Cam0840+1201 &  4.4 &  2.2 &  --- &  5.2 &286.0 & 12.5 &  --- & 21.9 & 16.0 &  3.4 &  7.1 &  3.2 &  2.7 &  1.1 &  --- &  1.8 &  --- & 16.2 &  7.3  &0.02 \\

 & $\pm$ 0.4& $\pm$ 0.3& ---&  $\pm$ 0.2& ---  & $\pm$ 0.2&  ---& $\pm$ 0.3& $\pm$  0.3& $\pm$  0.2& $\pm$ 0.3&  $\pm$  0.2& $\pm$ 0.2  & $\pm$ 0.7& ---& $\pm$ 0.2  & ---& $\pm$ 5.1& $\pm$ 1.3 & \\

Tol1924-416 & 4.9& 1.5&  1.1&  3.5&286.0 &  8.9 &  3.5 & 14.7 & 11.4 &  2.9 &  5.2 &  2.4 &  2.0 & --- & 1.4 & 1.8 &  --- & 11.7 & 4.3 &0.10 \\ 

 &  $\pm$  0.2&  $\pm$  0.2& $\pm$  0.2&  $\pm$  0.2&$\pm$  0.6  &  $\pm$  0.3&  $\pm$  0.1& $\pm$  0.1& $\pm$  0.1&  $\pm$  0.1&  $\pm$  0.1&  $\pm$  0.2& $\pm$  0.1  & ---&  $\pm$  0.2& $\pm$  0.3  &  ---& $\pm$  1.4&  $\pm$  0.7 & \\ 

Tol2019-405(NE) &  --- &  ---&  ---&  7.4&286.0& 14.9&  ---& 30.1& 24.3&  ---&  6.1&  3.2&  2.8&  ---&  ---&  ---&  ---& 29.4&  8.7 &0.10 \\

 &  ---&  ---&  ---&  $\pm$  1.3&$\pm$  2.0  & $\pm$  1.5&  ---& $\pm$  2.8& $\pm$  2.8&  ---&  $\pm$  1.2& $\pm$  0.9& $\pm$  1.0  & ---& ---& ---  & ---& $\pm$  4.6& $\pm$  2.5 & \\
 
Tol2019-405  &  ---&  ---&  ---&  7.7&286.0& 17.4&  ---& 36.0& 29.9&  ---&  6.9&  ---&  ---&  ---&  ---&  ---&  ---& 30.2&  8.3 &0.10 \\

 &  ---& ---& ---& $\pm$  1.6&$\pm$  2.6  & $\pm$  1.9& ---& $\pm$  2.3& $\pm$  2.7&  ---& $\pm$  1.6& ---& ---  &  ---& ---& ---  & ---& $\pm$  6.4&  $\pm$  1.9 & \\ 

Tol2138-397    &  ---&  ---&  ---&  ---&286.0&  ---&  ---& 14.8&  7.5&  ---&  ---&  5.4&  ---&  ---&  ---&  ---&  ---& 14.7&  --- &0.11 \\ 

&  ---&  ---&  ---&  ---&$\pm$  7.4  & ---& ---& $\pm$  0.8& $\pm$  0.5& ---& ---& $\pm$  0.4& ---  &  ---& ---& ---  & ---& $\pm$  2.8& --- & \\ 



Tol2146-391  & 23.2&  ---&  ---&  3.4&286.0&  5.5&  ---& 11.0&  7.5&  4.5&  8.5&  2.9&  2.6&  ---&  1.5&  2.4&  ---& 28.2$^{a}$&  ---  &0.12 \\

& $\pm$  1.0& ---&  ---&  $\pm$  0.2&$\pm$188.6  & $\pm$  0.3& ---& $\pm$  0.8& $\pm$  0.5& $\pm$  0.2& $\pm$  0.5&  $\pm$  0.1& $\pm$  0.1  &  ---&  $\pm$  0.3& $\pm$  0.4  & ---& $\pm$  2.7& --- & \\

MCG-05-52-065(E) & --- &  --- &  ---& 25.5&286.0& 59.1&  5.6& 67.4& 46.5&  ---&  8.4&  ---&  ---& ---&  ---&  ---&  ---& 31.3&  4.6 &0.44 \\ 

 &  ---&  ---& ---& $\pm$  2.8&$\pm$  4.1  & $\pm$  2.9& $\pm$  1.9& $\pm$  2.8& $\pm$  2.2& ---& $\pm$  1.3& ---& ---  & ---& ---& ---  & ---& $\pm$  3.6& $\pm$  1.4 & \\ 

MCG-05-52-065(W) &  ---&  ---&  ---& 24.1&286.0& 89.4&  ---& 88.8& 69.8&  ---&  ---&  ---&  ---&  ---&  ---&  ---&  ---& 29.5&  --- &0.44 \\ 

 &  ---& ---& ---& $\pm$  4.5&$\pm$  8.4  & $\pm$  6.7& ---& $\pm$  8.7& $\pm$  6.6& ---& ---& ---& ---  & ---& ---& ---  & ---& $\pm$  4.5& --- & \\


Tol2240-384 & 2.9 & --- & --- & --- &286.0 & 3.8 &  2.1 &  6.8 &  5.6 &  --- &  4.4 &  --- & --- & --- & --- & --- &  2.8 & 8.6 &  --- &0.36 \\ 

&  $\pm$  0.6&  ---&  ---&  ---&---  &  ---&  $\pm$  0.3&  $\pm$  0.4& $\pm$  0.3&  ---& $\pm$  0.1& ---&  ---  &  ---&  ---&  ---  &  $\pm$  1.7&  $\pm$  3.2&  --- & \\ 

UM160(E)      &  ---&  ---&  ---&  3.2 & 286.0 & 10.0&  4.8 & 24.8 & 17.1 &  ---&  5.1&  ---&  ---&  ---&  ---&  ---&  ---& 13.6&  --- &0.26 \\ 

&  ---&  ---&  ---& $\pm$  0.2& $\pm$  3.6  & $\pm$  0.9&  $\pm$  1.6& $\pm$  1.6& $\pm$  0.5&  ---& $\pm$  0.4& ---& ---  &  ---& ---& ---  & ---& $\pm$  2.7&  --- & \\

UM160(W)      &  ---&  ---&  ---&  --- &286.0 & 13.4 &  ---& 38.6 & 27.2 &  ---&  ---&  ---&  ---&  ---&  ---&  ---&  ---& 23.6 &  --- & 0.26 \\

   &  ---&  ---&  ---&  ---& $\pm$  1.6  & $\pm$  1.6&  ---& $\pm$  0.8& $\pm$  1.4& ---& ---& ---&  ---  &  ---&  ---&  ---  &  ---& $\pm$  4.9&  --- & \\

UM166   &  ---&  ---&  ---& 17.4&286.0& 58.3&  ---& 34.4& 31.7&  ---&  6.2&  ---&  ---&  ---&  ---&  ---&  ---& 54.1\footnote{We have quoted the flux of [SIII]$\lambda$9532~\AA~ instead of [SIII]$\lambda$9069~\AA~}&  --- &0.25 \\ 

&  ---&  ---&  ---& $\pm$  1.6&$\pm$  2.8  & $\pm$  1.7& ---& $\pm$  2.3& $\pm$  2.0& ---& $\pm$  1.3& ---& ---  & ---& ---& ---  & ---& $\pm$  5.8&  --- & \\

UM167 &  5.9  & --- &  1.4 & 36.6 &286.0 &105.6 &  2.5 & 30.2 & 29.8 &  2.8 &  4.0 &  3.0 &  2.4 & 0.8& 0.9 & 1.2&  2.3& 24.1&  4.1 &0.30 \\ 

&  $\pm$  0.3&  ---&  $\pm$  0.1& $\pm$  1.1&$\pm$  2.1  &$\pm$  0.6&  $\pm$  0.3& $\pm$  0.5& $\pm$  0.7& $\pm$  0.3& $\pm$  0.1& $\pm$  0.2& $\pm$  0.1  &  $\pm$  0.1&  $\pm$  0.2& $\pm$  0.2  &  $\pm$  0.6& $\pm$  2.8&  $\pm$  0.9 & \\ \hline

\end{tabular}
}
\vfill
\end{sidewaystable*}

The emission lines corresponding to the red spectra were measured
following the same procedure as in Kehrig et al. (2004). Once the
atmospheric correction was performed for each spectrum, we 
estimated the final error for each line flux by means of independent,
repeated measurements.

\begin{figure}
   \centering
    \includegraphics[bb=21 270 570 591,width=\columnwidth,clip]{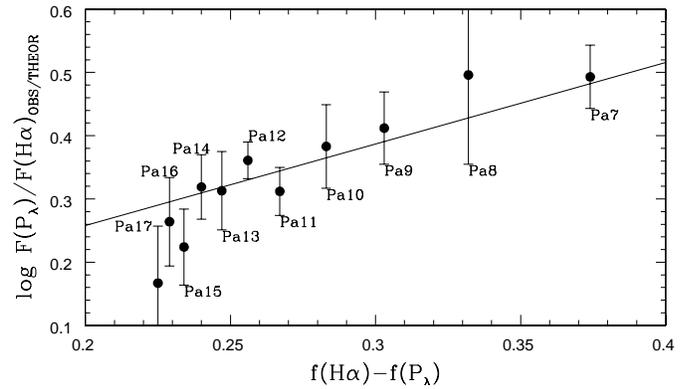}
      \caption{The ratio between observed and theoretical Paschen to H$\alpha$ flux, $P_{\lambda}$/H$\alpha$, versus the reddening function, f(H$\alpha$) - f(P$_{\lambda}$), for IIZw40. The line shows the least-square fit to the data.
              }
         \label{IIZw40fig}
   \end{figure}

We measured, for each galaxy, the main emission lines 
from the blue to the near-IR: [OII]$\lambda$3727; [OIII]$\lambda$4363; H$\beta$; [OIII]$\lambda$$\lambda$4959,5007; H$\alpha$;
[NII]$\lambda$$\lambda$6548,84; 
[SII]$\lambda$$\lambda$6717,31;
[SIII]$\lambda$$\lambda$$\lambda$6312,9069,9532; [ArIII]$\lambda$7136;
[OII]$\lambda$$\lambda$7320,30; Pa9 and Pa8, among others. The other Hydrogen
Paschen lines series, from Pa13 to Pa8, were detected in some galaxies
of our sample. Reddening-corrected line intensity ratios (applying Whitford's 1958 extinction law) normalized to
H$\beta$=100 are presented in Table~\ref{fluxes}, together with the
values of the reddening coefficient, C(H$\beta$), estimated using the H$\alpha$/H$\beta$ ratio from our blue spectra (Osterbrock 1989). Column (1) lists the
common names of the galaxies and, in those cases where the apertures
were centered on a secondary knot, there is an indication between
brackets for the position of the aperture (see Kehrig et al. 2004 for details). 
As is well known, IIZw40 presents high extinction (Baldwin et al. 1982). For this galaxy 
we measured the Paschen series from Pa17 to Pa8, thereby allowing a direct derivation of the reddening coefficient
from its red spectrum. 
In Fig.~\ref{IIZw40fig}, we illustrate for this galaxy the observed ratio of the fluxes of
each Paschen line to H$\alpha$, relative to its theoretical ratio for
case B recombination (Storey \& Hummer 1995),
[F(P$\lambda$)/F(H$\alpha$)]$_{obs}$/[F(P$\lambda$)/F(H$\alpha$)]$_{theo}$,
vs. the reddening function relative to H$\alpha$, f(H$\alpha$) -
f(P$\lambda$).

In order to check the reliability of the reduction process, we carried
out two tests.  Firstly, when the quality of the measurements allowed,
we compared the four brightest Paschen lines, normalized to H$\alpha$,
with the corresponding predictions for case B recombination (Storey \&
Hummer 1995). The corrected $P_{\lambda}$/H$\alpha$ values are found
to be consistent with the theoretical values, within the errors. In
the case of Pa8, we notice that the ratios are slightly above the
theoretical value. This is mainly due to the fact that the Pa8 and
[SIII]$\lambda$9532~\AA~lines are blended, making the Pa8 flux suffer
from some contamination by the [SIII] line. In the second test we
compared the ratio of the two near-IR [SIII] lines,
Q[SIII]=[SIII]$\lambda$9532/$\lambda$9069~\AA~, with the theoretical
ratio of 2.44 (Mendoza \& Zeippen 1982). The values of Q[SIII] are
consistent with the theoretical ratio to within the errors, although
many galaxies show Q[SIII] values slightly below the theoretical
ratio.  This effect could be due to two factors: (a) the telluric
absorption features were not totally removed and/or (b) the
[SIII]$\lambda$9532~\AA~line flux could be blended with the Pa8 line.

These two tests lead us to conclude that the correction for
atmospheric absorption, though not perfect, has provided generaly
satisfactory results for the purposes of this study.

\section{Results and Discussion}

\subsection{Empirical abundance indicators for our sample}

Commonly used strong line empirical abundance indicators are R$_{23}$
(Pagel et al. 1979; Edmunds \& Pagel 1984; McCall et al.  1985;
McGaugh 1991) and S$_{23(4)}$ (V\'{\i}lchez \& Esteban 1996; D\'{\i}az
\& P\'erez-Montero 2000; Oey \& Shields 2000; PM06).  Though widely
used, R$_{23}$ presents the drawback of having a double-valued
relation with oxygen abundance, creating an intrinsic uncertainty on
the derived O/H abundances. The turnover region of the relation
R$_{23}$ vs. O/H takes place for log R$_{23}$ $\gtrsim$ 0.9,
corresponding to 8.0 $\lesssim$ 12+log (O/H) $\lesssim$ 8.4. In this
region, R$_{23}$ is sensitive to ionization conditions but almost
insensitive to O/H. Most of the HII galaxies from our sample show
R$_{23}$ values within this ill- defined region, which is what we want to
explore.

The S$_{23}$ parameter introduced by V\'{\i}lchez \& Esteban (1996)
has been used as an O/H abundance calibrator in D\'{\i}az $\&$
P\'erez-Montero (2000) and P\'erez-Montero \& D\'{\i}az 2005
(hereinafter PMD05). It has also been demonstrated that S$_{23}$ is an
efficient S/H abundance calibrator in PM06. It presents several
advantages over R$_{23}$. First, it has a lower dependence on the
ionization parameter and remains single-valued up to metallicities
higher than solar, 12+log (O/H)$_{\odot}$ = 8.69 and 12+log
(S/H)$_{\odot}$ = 7.19 (Lodders 2003). Secondly, the sulfur emission
lines are less affected by reddening.  However, the spectral regions
around the red [SIII] lines are affected by atmospheric absorption.

The N2\footnote{N2=log ([NII]$\lambda$6584/H$\alpha$)} parameter has
also been proposed as an abundance indicator (Denicol\'o et al. 2002;
Van Zee et al. 1998). This parameter offers several advantages, because it
involves easily measurable lines that are available for a wide
redshift range (up to z $\sim$ 2.5). The N2 vs. O/H relation seems
monotonic and the [NII]/H$\alpha$ ratio does not depend on reddening
correction or flux calibration. The drawbacks are that the [NII] lines
can be affected by other excitation sources (see Van Zee et
al. 1998). In addition, N2 is sensitive to ionization conditions and
relative N/O abundance variations.

\begin{figure*}
\centering
\includegraphics[bb= 18 510 592 718,width=14cm,clip]{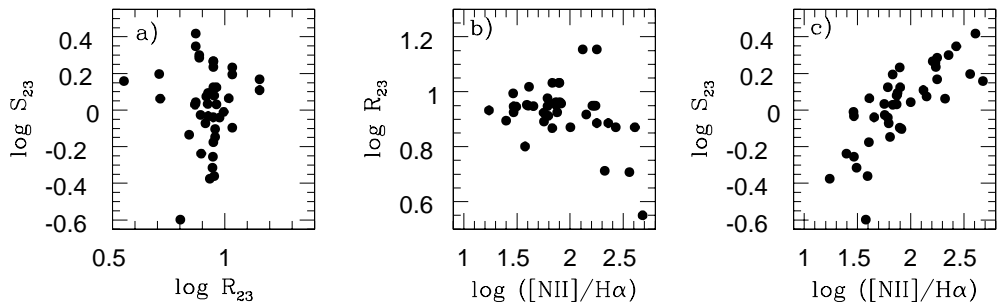}
\caption{The left panel presents the relation between S$_{23}$ and R$_{23}$; the middle and right panels  show 
 the relations between log (1.3{\it x}[NII]6584/H$\alpha$)
and the empirical indicators of abundances, R$_{23}$ and S$_{23}$ respectively, for
all galaxies of our sample.}

\label{S23_R23}
   \end{figure*}

\begin{figure*}
\centering
\includegraphics[bb=20 426 580 692,width=10cm,clip]{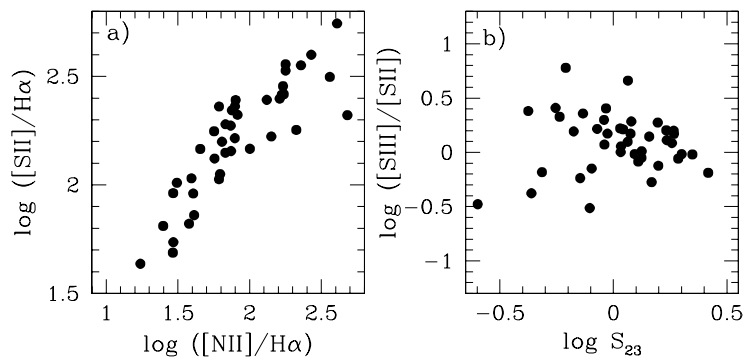}
\caption{The left panel shows the relation of log ([SII]6717,31/H$\alpha$) vs. log (1.3{\it x}[NII]6584/H$\alpha$); the right panel presents the relation between log ([SIII]/[SII]) and log S$_{23}$ 
for all galaxies of our sample.}

\label{NII_SII}
\end{figure*}

\begin{figure*}
\centering
\includegraphics[bb=20 426 580 692,width=10cm,clip]{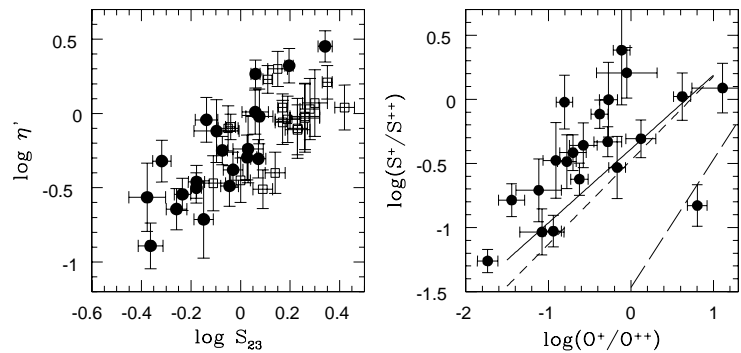}
\caption{The left panel shows the relation log $\eta$' vs. log S$_{23}$ for all galaxies in our sample. Full and empty symbols represent the objects with and without T$_{e}$[OIII], respectively. 
The right panel presents the relation between log (S$^{+}$/S$^{++}$) and log (O$^{+}$/O$^{++}$) for the galaxies with electron temperature; solid, short-dashed, and long-dashed lines show the loci for three series of photoionization models for T$_{eff}$=50kK, 40kK, and 30kK, respectively. See the text for details.} 
\label{cumulos}
\end{figure*}

Figure~\ref{S23_R23}a shows the relation between
S$_{23}$\footnote{The S$_{23}$ values were derived using the
sulfur emission lines quoted in Table~\ref{fluxes}} and
R$_{23}$\footnote{The R$_{23}$ values were calculated using the
oxygen emission lines from our blue spectra (Kehrig et al 2004).}  for the
galaxies of our sample. Although log R$_{23}$ values remain
approximately constant for most galaxies, log S$_{23}$ values present
a variation of approximately 0.8 dex. We can see that for galaxies in
the turn-over region of the relation between $R_{23}$ and O/H,
R$_{23}$ does not correlate with S$_{23}$. This fact is easily
understood since the relationship between $S_{23}$ and O/H is not
bivaluate in the metallicity range that we are interested in. Besides,
Figs.~\ref{S23_R23}b and c show that R$_{23}$ does not correlate
with [NII]/H$\alpha$, contrary to the behavior of
S$_{23}$. Therefore, for objects located in the ill-defined region of
R$_{23}$ vs. O/H, S$_{23}$ can be used to derive chemical abundances,
especially the S/H abundance.

\subsection{Ionization structure and the ionizing sources}

 In photoionized regions like the ones we consider here, the physical
 properties that determine line intensities are the luminosities and
 temperatures of the ionizing stars, the gas density, the optical
 thickness to the ionizing photons, and the chemical
 abundances. Because S$_{23}$ is a combination of strong line intensities, 
 it can be affected by several effects. Taking S$_{23}$ as an
 abundance indicator, we are not considering, to first order, the
 detailed effects produced by changes in the physical properties
 mentioned above. For this reason, it is important to check the
 sensitivity of S$_{23}$ to some of these properties.

The optical thickness to ionizing photons is the first to assess.  As
can be seen in Fig.~\ref{NII_SII}a, [NII]/H$\alpha$ and
[SII]/H$\alpha$ present a strong correlation, discarding density
boundary effects for the sample galaxies (see e.g. McCall et
al. 1985); this correlation implies a statistically significant
relation between N$^{+}$/N and S$^{+}$/S, as expected from standard
HII region models.

Ratios of line intensities of elements in different ionization stages,
such as [OIII]/[OII] or [SIII]/[SII], are sensitive to combinations of
the luminosity, the gas density and geometry, and the radiation
hardness; but they are insensitive to abundances at first order, as
they originate in the same element. Any variation with respect to
such line ratios indicates a sensitivity to these physical parameters,
though in a combination that might not be straightforward to derive.

Figure~\ref{NII_SII}b shows the dependence of S$_{23}$ on
[SIII]/[SII]\footnote{[SIII]/[SII] =
[SIII]$\lambda$$\lambda$9069,9532/[SII]$\lambda$$\lambda$6717,6731}. While
[NII]/H$\alpha$ shows a well-known dependence on the excitation degree
(e.g. McCall et al. 1985), the dependence is much weaker for S$_{23}$,
being mostly marginal. Despite the fact that S$_{23}$ possesses a narrower
dynamical range than [NII]/H$\alpha$, we consider it a better
abundance indicator for our sample than [NII]/H$\alpha$, since
S$_{23}$ does not show any strong dependence on the ionization
conditions.

Having a wide wavelength coverage has allowed us to 
study the properties of the ionizing sources in our sample of HII galaxies. 
This study could help to constrain the range of applicability of photoionization models 
and stellar atmospheres in order to fit the observations, thus improving 
our understanding of the mechanisms that heat the HII regions 
in HII galaxies (Stasi\'nska \& Schaerer 1999; Thuan \& Izotov 2005). 
A convenient hardness index is the parameter $\eta$' introduced by V\'{\i}lchez \&
Pagel (1988):

\[\eta' = \frac{[OII]\lambda\lambda3727,29/[OIII]\lambda\lambda4959,5007}{[SII]\lambda\lambda6717,31/[SIII]\lambda\lambda9069,9532} \]
 This parameter has been recommended as a criterion for effective
temperature of the ionizing star(s), T$_{eff}$, of HII regions, in the sense that softer ionizing spectra have higher values of $\eta$' (e.g. V\'{\i}lchez \&
Pagel 1988; Kennicutt et al. 2000). In Fig.~\ref{cumulos}a we present the behavior of
the parameter $\eta$' with respect to S$_{23}$ for our
sample. Overall, we can see that $\eta$' goes up with S$_{23}$, 
implying that the hardness of the ionizing spectra
increases with lower gaseous metallicity for our sample of HII galaxies (e.g. Bresolin \& Kennicutt 1999; Oey et al. 2000 and references therein; see also Mart\'{\i}n-Hernandez et al. 2002). 
A possible explanation for higher temperatures at lower
metallicities has been suggested by Massey et
al. (2004). They find that, for a range of stellar
atmosphere models, stars of early through mid-O types in a Magellanic
Cloud sample are 3000K-4000K hotter than their Galactic (metal-richer)
counterparts; and they attribute their higher temperatures to the minor
importance of wind emission, wind blanketing, and metal-line
blanketing at lower metallicities.

In Fig.~\ref{cumulos}b we present the relationship between the ionic
ratios S$^{+}$/S$^{++}$ and O$^{+}$/O$^{++}$ for the subset of HII
galaxies with electron temperature. In this figure we show the loci of
the average predictions of three sequences of single-star
photoionization models (computed with the photoionization code Cloudy 96; Ferland 2002), performed using CoStar model atmospheres
(Schaerer \& de Koter 1997) T$_{eff}$=50kK, 40kK
 and 30kK. Along each line, the
metallicities vary between $Z_{\odot}$/20 and $Z_{\odot}$/2, and the
ionization parameter changes from log U = -2 to log U = -3 (a detailed
description of the grids of the photo-ionization models used in this work
can be found in PMD05). According to these models, a large fraction of
the galaxies appear to harbor ionizing sources with spectra harder
than the spectrum produced by a 50kK effective temperature CoStar
atmosphere (Schaerer \& de Koter 1997). Kennicutt et al. (2000) have
found, for a sample of HII regions (in the Galaxy and Magellanic
Cloud), that empirically-based stellar-temperature indices present a
decrease in mean stellar temperature with increasing abundance. They
show, however, that the typical T$_{eff}$ for their HII regions are below
$\sim$ 55kK (at $Z_{\odot}$/5), in agreement with the model-based
results by Bresolin et al. (1999). Though any calibration of nebular
empirical parameters in terms of T$_{eff}$ should be a function of the
atmosphere and photoionization models used, it seems that T$_{eff}$ $\sim$ 55kK  represents a reasonable upper limit for the
effective temperature in HII regions in contrast to HII galaxies.
These findings suggest the existence of very hard spectral energy
distributions as ionizing sources in some HII
galaxies. Stasi\'nska \& Schaerer (1999), modelling the HII regions in
IZw18, argue that extra heating sources might well exist, in
addition to ionizing clusters, giving rise to large temperature
variations and enhancing the [OIII]$\lambda$4363 emission. I05 have
also invoked extra heating sources (i.e. X-ray ionizing sources) to explain
the high-ionization emission lines observed in some metal-poor
emission-line galaxies.  More observations, covering a wide range in
wavelength, as well as dedicated work using photoionization models for
evolving starbursts with a library of different ionizing spectra, are needed
to further investigate  the above suggestions.

\begin{figure}
\centering
 \includegraphics[width=9cm,clip]{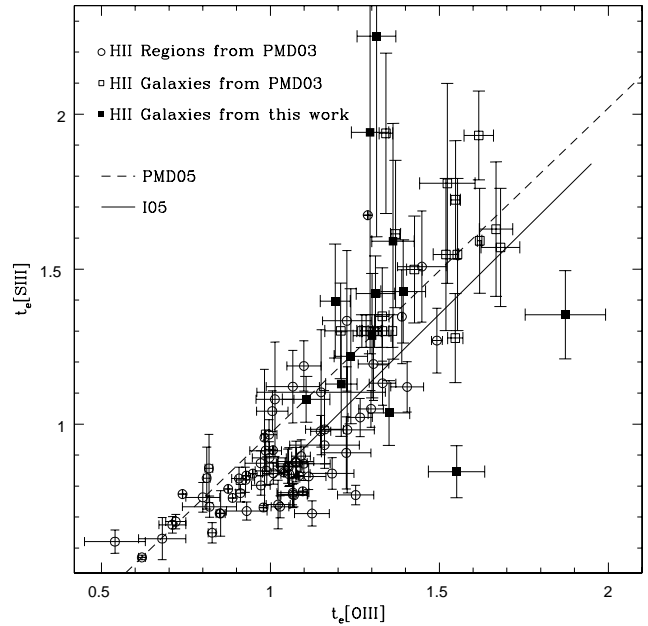}
        \caption{A comparison between the measured line temperatures
        of [OIII] and [SIII]. The electron temperatures, t$_{e}$[OIII] and t$_{e}$[SIII], are shown in units of 10$^{4}$K.
        The dashed and solid lines are the photoionization
        models'relation between these temperatures from PMD05 and
        I05, respectively.} 

\label{graf_temp}
\end{figure}

\begin{table*}
\caption{Physical properties and chemical abundances for the galaxies with S/H and O/H derived directly}  
\label{6gal}      
\centering
\renewcommand{\footnoterule}{}  
\begin{minipage}{\textwidth}
\begin{tabular}{lcccccccc}
\hline\hline 

Object & \multicolumn{3}{c}{IIZW40} & Tol0226-390 & Tol1924-416 & Tol0538-416  \\
       &G89\footnote{References - G89: Garnett (1989), PMD03: P\'erez-Montero \& D\'{\i}az (2003)} &PMD03$^{a}$ &This Work  &  &  &  &  & \\

\hline

$n_{e}$([SII])     &100. &290$\pm$60 &197$\pm$167   &221 $\pm$64 &78$\pm$28  &$\leq$ 100.\footnote{Low Density Limit}    \\

$T_{e}$([SIII])/10$^{4}$K  &1.35$\pm$0.12 &1.30$\pm$0.05  &1.04\footnote{$T_{e}$([SIII]) derived using [SIII]$\lambda$6312~\AA~ line from the red spectra}$\pm$0.10  &1.40$^{c}$$\pm$   0.18  &1.43$^{c}$$\pm$   0.17  &     1.29$^{c}$$\pm$   0.20  \\

$T_{e}$([OIII])/10$^{4}$K  &1.33$\pm$0.02  &1.34$\pm$0.03 &1.35$\pm$   0.06 &    1.19$\pm$   0.04 &    1.39$\pm$   0.07 &    1.30$\pm$   0.06 \\
12+log(S$^{+}$/H$^{+}$) &5.21 $\pm$0.09 &5.21$\pm$0.09  &5.23$\pm$   0.12   &    5.81$\pm$   0.07   &    5.50$\pm$   0.06   &    6.18$\pm$   0.04   \\
12+log(S$^{++}$/H$^{+}$) &6.00 $\pm$0.11 &5.99$\pm$0.04  &6.24$\pm$0.14   &6.17$\pm$0.16   &5.91$\pm$   0.13   &    6.20$\pm$   0.21   \\
12+log[(S$^{+}$ + S$^{++}$)/H$^{+}$] &6.07$\pm$0.11  &6.07$\pm$0.05  &    6.28$\pm$   0.14   &    6.33$\pm$   0.13   &    6.06$\pm$   0.11   &    6.49$\pm$   0.13   \\
ICF(S$^{+}$ + S$^{++}$) &1.95 &1.86  &    1.98    &    1.05    &    1.06   &    1.06   \\
12+log (S/H) &6.36$\pm$0.11 &6.34$\pm$0.05  &    6.53$\pm$   0.34   &    6.35$\pm$   0.14   &    6.08$\pm$   0.12   &    6.53$\pm$   0.14   \\
12+log(O$^{+}$/H$^{+}$) &6.95$\pm$0.07 &7.08$\pm$0.07  &    6.88$\pm$   0.26   &    7.37$\pm$   0.15   &    7.13$\pm$   0.14   &    6.99$\pm$   0.07   \\
12+log(O$^{++}$/H$^{+}$) &8.01$\pm$0.02 &8.03$\pm$0.08  &    7.96$\pm$   0.07   &    7.94$\pm$   0.07   &    7.83$\pm$   0.08   &    7.79$\pm$   0.07   \\
12+log (O/H) &8.05$\pm$0.02 &8.08$\pm$0.03  & 7.99$\pm$   0.09   &8.05$\pm$   0.09   &7.91$\pm$   0.09   &    7.86$\pm$   0.07   \\

\hline\hline

Object &  Cam0840+1201 &  CTS1008 &  UM238 &  Tol0104+388 &  UM306 &  UM307 \\ \hline

$n_{e}$([SII]) &$\leq$ 100.$^{b}$  &169$\pm$50 &$\leq$ 100.$^{b}$ &$\leq$ 100.$^{b}$  &111$\pm$109 &845$\pm$102      \\
 $T_{e}$([SIII])/10$^{4}$K  &     1.59$^{c}$$\pm$   0.38  &     1.22$^{c}$$\pm$   0.22  &     1.61\footnote{$T_{e}$([SIII]) derived from the relation $T_{e}$([SIII]) = 10500$T_{e}$([OIII]) - 800, see the text for details.}$\pm$   0.09  &     0.85\footnote{$T_{e}$([SIII]) derived using [SIII]$\lambda$6312~\AA~ line from the blue spectra}$\pm$   0.08  &  1.13$^{e}$$\pm$   0.17  &     1.08$^{e}$$\pm$   0.07  \\
 $T_{e}$([OIII])/10$^{4}$K  &    1.36$\pm$   0.06   &    1.24$\pm$   0.05   &    1.61$\pm$   0.09   &    1.55$\pm$   0.08   &    1.21$\pm$   0.05   &    1.11$\pm$   0.07   \\
12+log(S$^{+}$/H$^{+}$) &    5.52$\pm$   0.05   &    5.50$\pm$   0.11   &    5.54$\pm$   0.17   &    6.13$\pm$   0.03   &    5.83$\pm$   0.05   &    6.40$\pm$   0.17   \\
12+log(S$^{++}$/H$^{+}$) &    6.00$\pm$   0.29   &    6.20$\pm$   0.22   &    5.34$\pm$   0.09   &    6.96$\pm$   0.16   &    6.36$\pm$   0.24   &    6.31$\pm$   0.10   \\
12+log[(S$^{+}$ + S$^{++}$)/H$^{+}$] &    6.12$\pm$   0.24   &    6.28$\pm$   0.20   &    5.75$\pm$   0.14   &    7.02$\pm$   0.14   &    6.48$\pm$   0.20   &    6.66$\pm$   0.14   \\
ICF(S$^{+}$ + S$^{++}$) &    1.74   &    1.09   &    1.01   &    1.00   &    1.09   &    1.04   \\
12+log (S/H) &    6.36$\pm$   0.33   &    6.32$\pm$   0.22   &    5.76$\pm$   0.14   &    7.02$\pm$   0.18   &    6.52$\pm$   0.21   &    6.68$\pm$   0.14   \\
12+log(O$^{+}$/H$^{+}$) &    6.86$\pm$   0.14   &    6.90$\pm$   0.27   &    7.60$\pm$   0.36   &    8.33$\pm$   0.09   &    7.75$\pm$   0.07   &    8.68$\pm$   0.35   \\
12+log(O$^{++}$/H$^{+}$) &    7.77$\pm$   0.07   &    8.01$\pm$   0.07   &    7.65$\pm$   0.08   &    7.53$\pm$   0.08   &    7.92$\pm$   0.07   &    7.57$\pm$   0.11   \\
12+log (O/H) &    7.82$\pm$   0.08   &    8.05$\pm$   0.09   &    7.92$\pm$   0.23   &    8.40$\pm$   0.09   &    8.14$\pm$   0.07   &    8.71$\pm$   0.34   \\

\hline\hline

Object &  UM323 &  Tol0140+420 &  UM391 &  UM396 &  UM408 &  Tol0306+405 \\ \hline

$n_{e}$([SII]) &630:$^{f}$ &144:\footnote{Upper limit} &$\leq$ 100.$^{b}$ &$\leq$ 100.$^{b}$ &$\leq$ 100.$^{b}$ &220$\pm$64    \\
 $T_{e}$([SIII])/10$^{4}$K  &     1.35$^{e}$$\pm$   0.14  &     2.25$^{e}$$\pm$   0.65  &     1.09$^{e}$$\pm$   0.10  &     1.31$^{d}$$\pm$   0.06  &     1.71$^{d}$$\pm$   0.11  &     1.94$^{e}$$\pm$   0.71  \\
 $T_{e}$([OIII])/10$^{4}$K  &    1.87$\pm$   0.12   &    1.31$\pm$   0.06   &    1.12$\pm$   0.09   &    1.32$\pm$   0.06   &    1.71$\pm$   0.10   &    1.29$\pm$   0.06   \\
12+log(S$^{+}$/H$^{+}$) &    5.88$\pm$   0.13   &    5.83$\pm$   0.07   &    6.02$\pm$   0.05   &    5.30$\pm$   0.04   &    5.36$\pm$   0.05   &    5.94$\pm$   0.04   \\
12+log(S$^{++}$/H$^{+}$) &    5.86$\pm$   0.13   &    5.83$\pm$   0.28   &    6.33$\pm$   0.14   &    6.33$\pm$   0.12   &    5.47$\pm$   0.10   &    5.56$\pm$   0.43   \\
12+log[(S$^{+}$ + S$^{++}$)/H$^{+}$] &    6.17$\pm$   0.13   &    6.13$\pm$   0.19   &    6.50$\pm$   0.11   &    6.37$\pm$   0.11   &    5.72$\pm$   0.08   &    6.09$\pm$   0.19   \\
ICF(S$^{+}$ + S$^{++}$) &    1.02   &    1.03   &    1.02   &    1.78   &    1.01   &    1.01   \\
12+log (S/H) &    6.18$\pm$   0.13   &    6.16$\pm$   0.18   &    6.51$\pm$   0.11   &    6.61$\pm$   0.17   &    5.73$\pm$   0.08   &    6.13$\pm$   0.16   \\
12+log(O$^{+}$/H$^{+}$) &    7.99$\pm$   0.07   &    7.51$\pm$   0.08   &    7.49$\pm$   0.12   &    7.03$\pm$   0.07   &    7.18$\pm$   0.07   &    7.78$\pm$   0.07   \\
12+log(O$^{++}$/H$^{+}$) &    7.37$\pm$   0.08   &    7.78$\pm$   0.07   &    7.37$\pm$   0.14   &    7.97$\pm$   0.07   &    7.56$\pm$   0.08   &    7.90$\pm$   0.07   \\
12+log (O/H) &    8.08$\pm$   0.07   &    7.96$\pm$   0.08   &    7.74$\pm$   0.13   &    8.01$\pm$   0.07   &    7.71$\pm$   0.08   &    8.15$\pm$   0.07   \\

\hline\hline

Object &  Cam0357+3915 &  CTS1006 &  Tol2138+397 &  Tol2146+391 &  Tol2240+384 &  UM151\footnote{[OII]$\lambda$3727~\AA~ line could not be measured for UM151} \\ \hline
$n_{e}$([SII]) &$\leq$ 100.$^{b}$ &$\leq$ 100.$^{b}$ &$\leq$ 100.$^{b}$ &$\leq$ 100.$^{b}$ &132$\pm$80 &$\leq$ 100.$^{b}$      \\
 $T_{e}$([SIII])/10$^{4}$K  &     1.52$^{d}$$\pm$   0.08  &     1.42$^{e}$$\pm$   0.12  &     1.99$^{d}$$\pm$   0.14  &     1.68$^{d}$$\pm$   0.10  &     1.60$^{d}$$\pm$   0.09  &1.88$^{d}$$\pm$   0.12  \\
 $T_{e}$([OIII])/10$^{4}$K  &    1.52$\pm$   0.08   &    1.31$\pm$   0.06   &    1.97$\pm$   0.13   &    1.68$\pm$   0.10   &    1.60$\pm$   0.09   &    1.86$\pm$   0.12   \\
12+log(S$^{+}$/H$^{+}$) &    5.07$\pm$   0.07   &    5.70$\pm$   0.07   &    5.20$\pm$   0.04   &    4.93$\pm$   0.05   &    5.19$\pm$   0.05   &    5.77$\pm$   0.06   \\
12+log(S$^{++}$/H$^{+}$) &    5.86$\pm$   0.11   &    6.03$\pm$   0.09   &    5.82$\pm$   0.12   &    6.20$\pm$   0.08   &    5.68$\pm$   0.20   &    6.13$\pm$  0.13   \\
12+log[(S$^{+}$ + S$^{++}$)/H$^{+}$] &    5.93$\pm$   0.10   &    6.20$\pm$   0.09   &    5.91$\pm$   0.10   &    6.22$\pm$   0.07   &    5.80$\pm$   0.17   & 6.29$\pm$0.11   \\
ICF(S$^{+}$ + S$^{++}$) &    1.14   &    1.06  &    1.12   &    1.10   &    1.14   & 1.04    \\
12+log (S/H) &    5.98$\pm$   0.13   &    6.22$\pm$   0.09   &    5.96$\pm$   0.12   &    6.26$\pm$   0.09   &    5.86$\pm$   0.20   &6.31$\pm$0.12    \\
12+log(O$^{+}$/H$^{+}$) &6.33$\pm$0.14   &    7.54$\pm$   0.15   &    6.69$\pm$   0.07   &    5.88$\pm$   0.10   &    6.92$\pm$   0.08   & ---   \\
12+log(O$^{++}$/H$^{+}$) &7.78$\pm$   0.08   &7.82$\pm$   0.07   &    7.31$\pm$   0.08   &    7.61$\pm$   0.08   &    7.70$\pm$   0.08   & 7.05 $\pm$ 0.08  \\
12+log (O/H) &7.79$\pm$   0.08   &    8.00$\pm$   0.10   &    7.41$\pm$   0.08   &    7.62$\pm$   0.08   &    7.77$\pm$   0.08   & ---   \\

\hline
                  
\end{tabular}
\end{minipage}
\end{table*}

\subsection{Physical properties and chemical abundances}
\label{total_abund}

The physical properties and chemical abundances of the ionized gas
were calculated for these galaxies following the 5-level atom FIVEL
program (Shaw \& Dufour 1994) available in the task IONIC of the
STSDAS package. The final quoted errors in the derived quantities were
calculated by error propagation including errors in flux measurements,
atmospheric corrections, and temperatures. For the [SIII] lines we 
adopted the most recent atomic coefficients (Tayal \& Gupta 1999).

Electron densities were obtained from the
[SII]$\lambda$6717/$\lambda$6731~\AA~line ratio. We could derive the
electron temperature values of T$_{e}$[SIII], T$_{e}$[OIII],
T$_{e}$[OII], and T$_{e}$[SII] by combining the data from our blue
(Kehrig et al. 2004) and red spectra. Using the
[OIII]$\lambda$4363\AA/$\lambda\lambda$4959,5007~\AA~line ratio, we
derived the T$_{e}$[OIII] for 21 galaxies of the sample. The
T$_{e}$[SIII] was calculated from the
[SIII]$\lambda$6312/$\lambda\lambda$9069,9532~\AA~line ratio for 14
galaxies with a measurement of the [SIII]$\lambda$6312~\AA~ line. For
the 8 galaxies without any measurement of the
[SIII]$\lambda$6312~\AA~line and with T$_{e}$[OIII], a theoretical
relation between [OIII] and [SIII] electron temperatures (PMD05) was used:

\[T_{e}[SIII] = 10,500T_{e}[OIII]-800\]
In total we have 22 galaxies with a measurement of T$_{e}$[SIII]. 
In order to derive T$_{e}$[SIII] temperature and S$^{++}$ abundances,
whenever possible we used the [SIII]$\lambda$9069~\AA~fluxes, given
that the flux of [SIII]$\lambda$9532~\AA~seems to be affected by partial blending
of the P8 line. However, when the [SIII]$\lambda$9069~\AA~line falls inside the
absorption band head due to the redshift of the galaxy, the flux of the
[SIII]$\lambda$9532~\AA~line was used if this line was not close to
the 1$\mu$m limit, where the flux
calibration becomes highly uncertain.

Regarding [SII] temperatures, for those objects without the [SII]
auroral line at $\lambda$4068~\AA~we took the approximation
T$_{e}$[SII] $\approx$ T$_{e}$[OII] as valid. We could derive
T$_{e}$[OII] using the [OII]$\lambda$ $\lambda$3727/7325~\AA~line
ratio for 16 objects of the sample. For the rest of the objects not
presenting any auroral line in the low excitation zone, we used the
model-predicted relations between T$_{e}$[OII] and T$_{e}$[OIII] found
in PMD03, which explicitly take the dependence of T$_{e}$[OII] on
electron density into account. In most cases the agreement between our
line-intensity measurements in the blue and in the red spectra is
good; we thus have adopted the values with the smaller observational
errors to derive line temperatures. Otherwise, for T$_{e}$[SIII] and
T$_{e}$[SII], we used the line intensities corresponding to the red
spectra.

The relationship between both temperatures, T$_{e}$[OIII] and
T$_{e}$[SIII], is shown in Fig.~\ref{graf_temp} for all our galaxies with
electron temperature and the sample of HII galaxies and HII regions compiled in
PMD03, together with photoionization model relations.
We note that there are two behaviors. While most HII
 regions show lower T$_{e}$[SIII] values than the ones provided by the
photoionization models relations, many HII galaxies present higher T$_{e}$[SIII] values
 than the ones predicted by the models.  The same trend can be
noticed for the sample of metal-poor emission-line galaxies in I05 (their Fig.4), 
for a range of T$_{e}$[OIII] from 1x10$^{4}$K to
2.0x10$^{4}$K. This fact suggests that HII regions and HII
galaxies probably present different spatial temperature structures.

In order to compute the total sulfur abundances, we need to
evaluate the corresponding ICF. A detailed study of the ICF scheme for
sulfur is described in PM06. According to this work, for the objects with log ([SIII]/[SII])
$\geq$ 0.4, we made use of the formula of Barker with $\alpha$ =2.5
(Barker 1980). For the rest of the objects in the sample, we used
the predictions of photoionization models for CoStar atmospheres (see
Fig.3 in PM06). These predictions indicate rather low values of the
ICF, independent of the ionizing effective temperature of the models.

With regard to the oxygen ICF, a small fraction of O/H is expected to
be in the form of O$^{3+}$ ion in the high-excitation HII regions when
the HeII$\lambda$4686 emission line is detected. We have a measurement
of the HeII$\lambda$4686 emission line in 6 galaxies of our
sample. According to the photoionization models from Stasi\'nska \&
Izotov (2003), the O$^{3+}$/O can be on the order of 1$\%$ only in the
highest-excitation HII regions [O$^{+}$/(O$^{+}$ + O$^{2+}$) $\leq$
0.1]; therefore, taking our abundance results into account, we assumed
that this correction is negligible in our sample.

Physical conditions, chemical abundances, and ICFs of sulfur for the
galaxies with a measurement of the T$_{e}$[SIII] are quoted in
Table~\ref{6gal}. From this Table we can see that there are six
galaxies with 12+log (O/H) varying between 7.4 and 7.8. These objects
are among the galaxies with very low metallicity.  For the objects
without T$_{e}$[SIII], we used the strong line calibration of the total
S/H abundance as a function of S$_{23}$, presented by PM06, to derive
the total S/H abundance.

\begin{figure}
\centering
\includegraphics[width=7.0cm,clip]{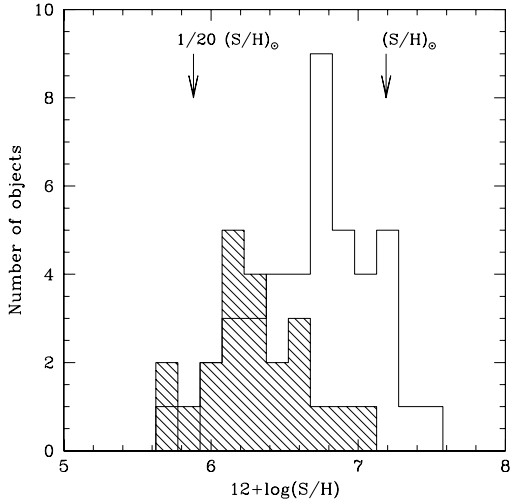}
\caption{\footnotesize The distribution of total sulfur abundance for our
sample of galaxies. The dashed and empty histograms show the number of galaxies
with S/H derived from $T_{e}$[SIII] (see Table~\ref{6gal}) and obtained from the S$_{23}$ calibration, respectively (see the text for details).}

\label{hist}
\end{figure}

\begin{figure}
\centering
\includegraphics[bb=21 260 570 591,width=\columnwidth,clip]{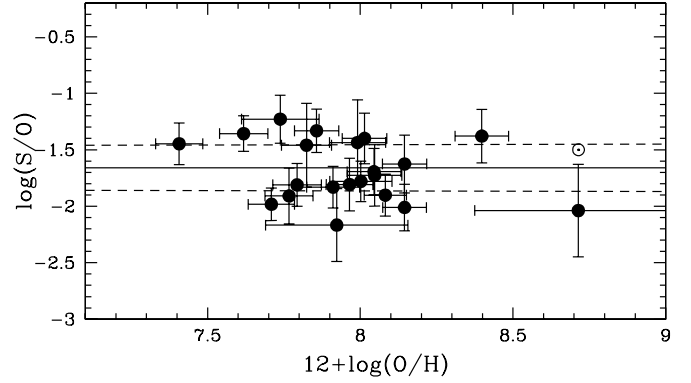}
\caption{\footnotesize
The observed sulfur-to-oxygen abundance ratio for the subset of
galaxies of the sample with $T_{e}$[OIII] plotted as a function of the oxygen abundance. The solar value is shown. 
The dashed lines
are +/- (1$\sigma$) of the mean as shown by the continuous
line.}
\label{S_H_O_H}
\end{figure}

The only galaxy of our sample for which we can compare the S/H
abundance derived in this work with previous S/H abundance
determinations in the literature is IIZw40. This galaxy has been
observed by G89 and PMD03. In Table~\ref{6gal} we present the results
for IIZw40 obtained by the three works. In order to minimize possible
reddening corrections effects in the abundance calculation for this
galaxy, we referred the flux of each sulfur line we measured to a
nearby hydrogen line. In the case of 12+log (S$^{+}$/H$^{+}$) ionic
abundance the three values are close to each other; the value of the
12+log (S$^{++}$/H$^{+}$) ionic abundance, derived in this work is
higher than the previous ones by up to some 0.2 dex. We believe that
this fact could be the result of our systematic absorption correction
procedure.

Figure \ref{hist} shows the distribution of sulfur abundance derived
for our sample of HII galaxies. The empty and dashed histograms
represent the distribution of S/H derived with S$_{23}$ for all the
objects and obtained from T$_{e}$[SIII], respectively. Most of the
galaxies present total S/H abundance values that are between 1/20
solar to solar\footnote{12+log (S/H)$_{\odot}$ = 7.19 $\pm$ 0.04
(Lodders 2003)}. This is an expected behavior since our sample is
composed mainly of low-luminosity galaxies. Besides, we note that the
dashed histogram peak corresponds to total S/H abundance value lower
than the S/H maximum of the empty histogram. It suggests that, in order
to know the overall metallicity distribution of a sample of galaxies,
it would be worth making use of an efficient empirical abundance
indicator. Hoyos $\&$ D{\'{\i}}az (2006) found a similar result by
studying the O/H abundance for a sample of HII galaxies.

The abundances obtained in this work allow us to study the dependence
of S/H as a function of O/H in low metallicity environments.  In
Fig.~\ref{S_H_O_H} we show the relationship between the S/O abundance
ratio and total O/H abundance for the subset of galaxies with
T$_{e}$[OIII] and T$_{e}$[SIII]. The value of the sigma weighted mean
for log(S/O) is -1.68$\pm$0.20 dex.  The galaxy with nearly solar
metallicity (UM307) is classified as an SABd from HYPERLEDA
database\footnote{leda.univ-lyon1.fr (Paturel 2003)}.

Evaluating the contribution of
all observational errors to the derivation of these abundances, we can conclude at
this level of uncertainty  that there is no statistical
evidence of any systematic variation of S/O with O/H for this range
of abundances. Therefore, our results agree with a constant 
S/O ratio and lower (1$\sigma$) than the solar ratio for this type of emission-line object.
This result indicates that sulfur and oxygen appear to be produced by the same massive stars,
as expected by current nucleosynthesis prescriptions (see Pagel 1997 and references therein).
In recent works, I05 and PM06 indicate that
HII galaxy data are consistent with a constant S/O ratio, but somewhat lower
than the current solar ratio. Regarding disk HII regions, the
dispersion in S/O appears much larger and the assumptions of a constant S/O
is questionable there. These results suggest that the
assumption that the S/O ratio is constant at all abundances remains
controversial (e.g Bresolin et al. 2004) and should
be explored further, particularly at the not very well-known metallicity
ends: extremely metal deficient HII galaxies (i.e. very low O/H) and
HII regions in the inner disk of galaxies (i.e. metal rich central
parts with highest O/H).

\section{Summary}

In this work we have performed a long-slit spectroscopic study of
a sample of 34 HII galaxies observed in the blue and near-IR ranges
(3700$\AA$-1$\mu$m). The red spectra were carefully corrected for the effects of the
telluric atmospheric absorption. Measurements of the nebular [SIII] lines at
$\lambda$$\lambda$9069,9532 were obtained for all objects.

Whenever possible we derived values of T$_{e}$[SIII], T$_{e}$[OIII],
T$_{e}$[OII], and T$_{e}$[SII] by combining our data in the red with
our data in the blue. Regarding T$_{e}$[SIII], most of the observed
HII galaxies show values that are slightly higher than those predicted
from T$_{e}$[OIII] by photoionization models. This effect can be
especially important for the high-excitation objects.

 We derived the total S/H abundance for the 34 objects in the
 sample. Total S/H abundance was calculated
 directly using the electron temperature T$_{e}$[SIII] in 22 HII galaxies, 
 for which the O/H abundance was obtained directly from the
 observations using T$_{e}$[OIII]. For the rest of the objects total
 S/H abundances were computed using the empirical abundance indicator
 S$_{23}$.

A comparative study was performed on the reliability of S$_{23}$
and R$_{23}$ as abundance indicators. No systematic variation in derived S/H
with the excitation degree of the HII regions was found. That means
that S$_{23}$ is not sensitive to ionization effects, at first order,
making it a robust empirical abundance indicator.

The  comparison between $\eta$' and S$_{23}$ parameters for our
sample indicates that harder ionizing spectra are found in the HII
galaxies with lower gaseous metallicity. 
Comparing the ionic ratios O$^{+}$/O$^{++}$ and
S$^{+}$/S$^{++}$ with the predictions of single-star photoionization models, we note
that a large fraction of galaxies in our sample are probably ionized
by very hard spectra. This result points out that extra heating sources might
exist, as has been suggested by recent works (Stasi\'nska \& Schaerer 1999; I05).

Finally, we presented a study of the abundance of S/H as a function of O/H
in low-metallicity environments. Our data, together
with other studies of S/H based upon the near-IR [SIII] lines, are
consistent with the conclusion that S/O remains constant as O/H varies
among the sample of HII galaxies. The scatter in S/O 
(due mainly to observational errors) is still large to constrain the
degree of variation in S/O over
the whole O/H abundance range. 
The assumption that the S/O ratio remains
constant at all abundances is still an open question that should be
explored further.

\begin{acknowledgements}
   
  C.K wishes to thank the Conselho Nacional de Desenvolvimento
  Cient\'{\i}fico e Tecnol\'{o}gico (CNPq-Brasil) for a grant and the
  Consejo Superior de Investigaciones Cient\'{\i}ficas (CSIC-Spain)
  for an I3P fellowship. We thank the referee for useful
  suggestions. We thank H.Plana for carrying out part of the
  spectroscopic observations. We also thank E.P\'erez,
  R.M.Gonz\'alez-Delgado and D. Reverte Pay\'a for their help in the
  initial stages of this project, and to Jorge Iglesias-P\'aramo for
  his fruitful comments and careful reading of the manuscript. This
  research was partially funded by project AYA2004-08260-C03-02 of the
  Spanish PNAYA.

\end{acknowledgements}

\appendix

\end{document}